\documentclass[10pt,twocolumn,twoside]{IEEEtran}
\usepackage{slashbox}
\usepackage{graphicx}
\usepackage{subfigure}
\usepackage{color}
\usepackage{multirow}
\usepackage{amsmath,amssymb,epsfig,theorem,url,cite,bm}
\usepackage{wrapfig }
\usepackage{soul}
\usepackage{multicol}
\usepackage{cuted}
\usepackage{tablefootnote}

\begin{document}

\title{Estimating Uniqueness of I-Vector Based Representation of Human Voice}

\author{Sinan E. Tandogan, H\"usrev~Taha~Sencar
\thanks{
S. E.~Tandogan is with TOBB ETU, Ankara, Turkey.
H. T. Sencar is with Qatar Computing Research Institute, HBKU, Qatar and TOBB ETU, Ankara, Turkey. 
(e-mail: {\em hsencar}@hbku.edu.qa)
}
}

\maketitle

\begin{abstract}

We study the individuality of the human voice with respect to a widely used feature representation of speech utterances, namely, the i-vector model. 
As a first step toward this goal, we compare and contrast uniqueness measures proposed for different biometric modalities.  
Then, we introduce a new uniqueness measure that evaluates the entropy of i-vectors while taking into account speaker level variations.
\textcolor{black}{
Our measure operates in the discrete feature space and relies on accurate estimation of the distribution of i-vectors.
Therefore, i-vectors are quantized while ensuring that both the quantized and original representations yield similar speaker verification performance.}
\textcolor{black}{Uniqueness} estimates are obtained from \textcolor{black}{two newly generated datasets and the public VoxCeleb dataset.}
The first custom dataset \textcolor{black}{contains} more than one and a half million speech samples of 20,741 speakers obtained from TEDx Talks videos.
The second one includes over twenty one thousand speech samples from 1,595 actors that are extracted from movie dialogues.
Using this data, we analyzed how several factors, such as the number of speakers, number of samples per speaker, sample durations, and diversity of utterances affect \textcolor{black}{uniqueness} estimates.   
Most notably, we determine that the discretization of i-vectors does not cause a reduction in speaker recognition performance. 
Our results show that the degree of \textcolor{black}{distinctiveness} offered by i-vector-based representation may reach 43-70 bits considering 5-second long speech samples;
however, under less constrained variations in speech, uniqueness estimates are found to reduce by around 30 bits.  
We also find that doubling the sample duration increases the distinctiveness of the i-vector representation by around 20 bits.

\end{abstract}

\vspace{-0.2cm}
\section{Introduction}
Biometric solutions have increasingly become a key component of security systems that govern everyday processes of private and public life. 
Today, most smartphones utilize at least one of fingerprint, facial, or iris recognition as the main method for user verification. 
Behavioral biometrics are at the core of enterprise information systems to detect anomalies in user interactions and to provide initial as well as continuous authentication.
This widespread adoption of biometrics for implementing authentication and user identification is evidently due to 
the fact that biometric traits are unique to the individual, that they are sufficiently invariant, and that they can easily be captured and processed with minimal user intervention.

Among all types of biometric technologies, there is a more visible growth in the use of voice-based authentication and identification systems. 
This is in part due to well-established expertise in audio signal representation and processing \textcolor{black}{and the increase in available computational power in many systems, especially in embedded systems.}
But more importantly, it is due to various application contexts that involve speech-based user interactions where voice as a biometric modality is readily available.
For example, in call center environments verifying the identity of a caller through voice offers a more convenient and cost-effective alternative to conventional knowledge-based authentication where the agent asks the caller a series of questions. 
Moreover, for embedded sensor-enabled devices such as wearables that have limited user interface there are only a few alternatives to voice for authentication.
With the pervasive availability of voice assistants in everyday devices, such as smartphones and IoT devices, 
and the potential integration of voice technology within many applications, speaker verification capabilities will likely become even more important. 

At its core, the use of a \textcolor{black}{biometric voice representation}, or any other biometric modality for the general case, as part of security systems rests on the uniqueness and individuality of \textcolor{black}{such biometric identifiers, which may refer to both biological and behavioral attributes.} 
\textcolor{black}{However, acquisition of biometric data through a lossy and error-prone process 
and the variability induced by several environmental factors effectively diminishes discernible characteristics, thereby limiting the biometric information content.
The fact that the representation for the captured voice samples of a speaker may not be unique essentially introduces a vulnerability that can be exploited by attackers.}

Voice authentication systems operate by constructing a model of each user's voice. 
With these systems, user verification can be performed in a text dependent or independent manner by evaluating the match of a voice model estimated from test samples to a more reliable model created during user enrollment. 
One possible attack against these system\textcolor{black}{s} includes the use of voice synthesis and conversion methods to generate the desired text in a voice that will produce a match with the target speaker, {\em i.e.,} voice-spoofing attacks, 
\cite{ergunay2015vulnerability, turner2019attacking}. 
\textcolor{black}{No matter how it is performed, such an attack requires access to voice samples of the target user of durations varying from a couple of minutes to several hours \cite{khodabakhsh2017spoofing, baljekar2018speech, Derrick2019}.}

Since launching such an attack at scale requires a large amount of samples from users, it is not practically feasible \textcolor{black}{most of the time}.  
\textcolor{black}{A fraudster, however, can alternatively use speech samples of other speakers with a similar voice to spoof the voice of a target speaker.}
For this purpose, the attacker can create a dictionary that spans a diverse range of voices, along with sufficient samples, and can quickly identify voice samples most similar to a given query utterance to be used during spoofing.
Unlike many other biometric modalities, it is relatively easy for fraudsters to capture voice for a large population. 
Not only are large numbers of voice samples available online, one can also acquire voice samples by systematically calling phone numbers. 
With robocalling and VoIP technology, such calls can be made at scale at very low cost.
Hence viability of this attack depends largely on the difficulty of creating a voice dictionary of sufficient size to be in the order of unique voice models. 
This ultimately calls for evaluation of the uniqueness of users' voice models, a task which translates into an inquiry into the information content of a biometric modality.

Despite significant research in representation and matching of biometric data, the uniqueness and individuality of biometric modalities have not been studied very thoroughly. 
To date, a number of studies have been undertaken to determine the amount of information contained in a biometric modality including voice, iris, face, and fingerprints. However, measurements provided by different publications vary significantly.
In the case of voice, individuality estimates range between 14 bits \cite{IEEEexample:yang2014} and 120 bits or higher depending on the length of speech samples used for measurements \cite{IEEEexample:entropy2015}. That is, the number of unique human voices is in the order of $2^{14}$ or $2^{120}$. 
Further, a similar pattern has been observed across other modalities, albeit to different extents \cite{sutcu2013biometric}. 
For example, a \textcolor{black}{single} fingerprint, the most mature and well established modality, is estimated to have an entropy of between 12.7 \cite{IEEEexample:takahashi2010} and 55 \cite{young2013entropy} bits.
Similarly, estimates of the uniqueness of biometric face images vary between 12.6 \cite{IEEEexample:takahashi2010} and 55.6 \cite{adler2009towards} bits.
Finally, iris as a biometric modality, which is considered to be the most reliable of all, is estimated to offer 249 \cite{IEEEexample:daugman2003} to 288 \cite{youmaran2012measuring} bits of information.
Variations up to an order of magnitude (in the exponent) difference in the above estimates indicate the need for a more systematic evaluation.  

In fact, three factors play an important role in the discrepancy between reported results on distinctiveness of biometric modalities. 
Biometric data is almost always \textcolor{black}{mapped} into a feature space. 
Hence, an important source of variability is the choice of biometric feature representation. 
Since compact representation of a continuous variable can only be achieved at the expense of information loss, different representations of a biometric modality yield varying  discriminative power. 
The other important factor contributing to difficulty in establishing the uniqueness of a biometric modality is the \textcolor{black}{measurement} of biometric information. 
The inherent variability of a biometric modality combined with the measurement noise and the complexity of modeling high-dimensional feature representations hinder analytical tractability significantly. 
As a result, it becomes difficult to directly utilize the concept of entropy and alternative definitions have been adopted to estimate the inherent entropy.
The last source of variability concerns the dataset used for modeling and measurements. 
Essentially, the accuracy of estimates depends on how well a given dataset reflects the biometric diversity of users and the overall biometric variability exhibited by users. 
However, producing reliable, comprehensive public datasets is a very challenging task due to increasing privacy concerns.

To help close this gap, in this work, we study the problem of measuring the individuality of a biometric modality in the context of voice biometrics.
Towards this goal, our work brings all approaches to measuring distinctiveness of biometric modalities together and evaluates their strengths and weaknesses from the standpoint of generalizing these measures.
Our approach to estimation of uniqueness differs from existing ones as it computes the entropy of speaker i-vectors while taking into account within-speaker 
variability and builds on a mutual information-based formulation. 
Measurements are performed on a widely used feature representation in speaker recognition, namely, the i-vector representation.
To be able to estimate the underlying feature distribution as accurately as possible, speaker i-vectors are quantized, 
and performance implications of operating in a discrete feature space are investigated.
To ensure that the uniqueness estimates obtained using quantized i-vectors are reflective of the original, non-quantized i-vectors, the speaker verification performances yielded by the two representations are set to similar values.

To evaluate and compare our approach, we created two distinct benchmark datasets in addition to the public VoxCeleb dataset. 
One of these datasets includes voice samples of close to \textcolor{black}{twenty-one} thousand speakers obtained from audio tracks of TEDx Talk videos, and the other
includes samples of 1,595 actors extracted from dialogues in 249 movies.
The former dataset is mainly used to quantify how uniqueness estimates vary depending on number of speakers as well as the amount of speech samples available from each speaker, whereas the latter is used to determine to what degree the true variability intrinsic to a speaker's voice affects estimates.
Overall, the data used in our experiments constitute the most comprehensive set used by a study of similar nature in its effort to better incorporate \textcolor{black}{between}-speaker and within-speaker variability.

Our paper is organized as follows.
In Section \ref{sec:ses}, we start by reviewing the work done in the field of  speaker recognition and speaker verification with an emphasis on voice models proposed for speaker representation. 
This is followed by a qualitative description of approaches proposed for measuring individuality of a variety of biometric modalities and a discussion of their applicability to voice biometrics in Section \ref{sec:method_entropy}. 
The details of our uniqueness estimation method are given in Section \ref{sec:method}. 
\textcolor{black}{The datasets used in experiments along with details about the generation of the two new datasets are described in Section \ref{sec:dataset}. }
The \textcolor{black}{evaluation protocol}, results of analyses and uniqueness estimates obtained considering a variety of settings are provided in Section \ref{sec:results}.
We conclude the paper with a discussion of our findings in Section \ref{sec:conclusion}.

\section{Voice Models}
\label{sec:ses}
\textcolor{black}{
Speech signals have most commonly been represented by the so-called Mel-Frequency Cepstral Coefficients (MFCCs) \cite{logan2000mel}.
Although recently deep learning techniques have also been proposed as a means to obtain alternative representations \cite{ravanelli2018speaker},  
the majority of speaker recognition systems today deploy MFCC-based feature representations.}
MFCCs provide spectral energy measurements over short-term frames of a speech signal with each measurement involving a vector of 10-20 coefficients. 
These coefficients in essence capture unique spectral characteristics of a speaker's voice and the manner with which a speaker articulates different sounds in the language. 
\textcolor{black}{To better capture the spectral dynamics of a speaker, the MFCCs obtained from each frame are further augmented by the first-order and second-order derivatives of the coefficients.}

\textcolor{black}{Several speaker modeling approaches are proposed for text-independent speaker recognition \cite{kinnunen2010overview}.
Among these, the Gaussian Mixture Model (GMM) has been the most effective \cite{el2017text}.}
When using a GMM, speakers are characterized by modeling the distribution of their MFCC vectors by a mixture of Gaussians \cite{IEEEexample:gmm}.
\textcolor{black}{Hence, each speaker model is represented by a set of GMM parameters that include the weights of a speaker's Gaussian component, the  mean vector and the covariance matrix.}
In addition to speaker-specific models, a Universal Background Model (UBM) is created which is a similarly generated model using speech samples of a large set of speakers to represent general, speaker-independent feature characteristics.
The speaker models and the UBM are used together to perform speaker verification \cite{IEEEexample:gmmubm}.  
In the resulting GMM-UBM system, a verification decision about an unknown speech sample is made through a likelihood ratio test which evaluates the degree of match
between the known speaker model and the UBM.

This initial system is later further improved by a focus on better modeling of speaker related variations while compensating for undesired variabilities.
With this objective, Campbell {\em et al.} introduced the concept of GMM mean supervector by stacking the mean vectors of each GMM component in a high-dimensional 
vector \cite{IEEEexample:campbell2006Supervector}.
Subsequent research focused on decomposing this mean supervector into a sum of speaker- and channel-dependent components.
Kenny {\em et al.} \cite{IEEEexample:JFA2006} showed that joint factor analysis can be effectively used for modeling these components.
Dehak {\em et al.} \cite{IEEEexample:ivector2011}, introduced an alternative factor analysis model which resulted in a representation that is widely used by 
various state-of-the-art speaker recognition systems. 
In this framework, the GMM mean supervector associated with a speaker's utterance is represented in a total variability space that contains both the channel 
and speaker variability, and the variability in the supervectors is compactly represented in terms of what is known as an identity vector (i-vector). 
An i-vector is essentially a fixed-dimensional representation of a speech utterance obtained by transforming the associated mean supervector using a  total variability matrix.
The i-vectors are typically projected into a linear discriminant analysis (LDA) space and post-processed to further improve their distinctiveness.

Although there are several methods for i-vector based speaker modeling and comparison, Gaussian Probabilistic Linear Discriminant Analysis (GPLDA) 
is the commonly employed one \cite{IEEEexample:gpdla2011}.
The GPLDA approach essentially divides the i-vector space further into speaker and session variability subspaces such that the speaker-specific component
assumes a standard normal distribution.
For speaker verification, a hypothesis test is performed to determine whether the underlying speaker specific component for the test i-vector is the same as those estimated from speaker i-vectors obtained during modeling.
This is realized by computing a log-likelihood ratio based score between the test i-vector and the speaker models.

With the advent of deep learning, more recently, deep neural network (DNN) architectures have also been used to build speaker models.
This approach effectively utilizes the outputs of a layer of a DNN as feature vectors.  
The most successful of such proposals is the so-called x-vector representation which incorporated the idea of data augmentation to improve the robustness of DNN embeddings obtained at the last fully-connected layer~\cite{snyder2018x-vector}.
Experimental evaluations show that use of neural network-based speaker embeddings improve speaker recognition performance.

In this work, we focus on measuring the uniqueness of i-vector representation of speech as it has been used widely in speaker verification systems.
Although neural speaker embeddings are currently more prevalent with several variants, i-vector embeddings are still 
competitive for longer utterance lengths \cite{snyder2020x}.
Moreover, it has been demonstrated that both in the text-independent \cite{kellyvectors} and text-dependent \cite{sarkar2020bottleneck} test scenarios the fusion of i-vector with newer embeddings improves the performance, indicating that these representations have also a complementary nature.
Regardless of the choice of speaker embedding, however, an important consideration is that they all couple speaker, channel, and session variabilities in the resulting model representation of the speech.
That is, the feature representation contains more information than just speaker information. 
Although it is possible to separate the speaker's variable when i-vector distribution is modeled with PLDA, this is shown to be very complicated for a very large number of users \cite{kinnunen2017non}.

\section{Uniqueness Estimation Approaches}
\label{sec:method_entropy}
The amount of discriminatory information present in a biometric modality has long been a focus of research. 
Early work mainly used the probability of false biometric matching, {\em i.e.} matching a given biometric sample to any other sample by chance for a single user verification attempt, as a measure for estimating individuality \cite{pankanti2002individuality}. 
Setting up a duality with password guessing attacks, O'Gorman \cite{o2003comparing} argued that the sample space of a biometric modality, defined as the valid range of values that can be taken by biometric features, can be used to estimate an upper bound on the individuality of a modality. 
Accordingly, the effective sample space is measured by the inverse of false matching probability which is then mapped to maximal entropy of a modality under the assumption of uniform distribution of sample values. 
In line with this thinking, Dass {\em et al.} \cite{zhu2007statistical} focused on deriving an expression to estimate the probability of a false correspondence between minutiae features of two arbitrary fingerprints. 
This is realized by modeling the distribution of biometric features and using the resulting models to generate random biometric samples needed to calculate the random correspondence probabilities.

Subsequent approaches to quantifying the amount of information available in different biometric modalities adopted alternative definitions that are more focused on modeling between-subject and within-subject variability of biometric features. 
Below, we provide a brief overview of these approaches, discuss their theoretical underpinnings, and evaluate their applicability to measuring distinguishability offered by the voice as a biometric modality.

\vspace{-0.3cm}
\subsection{\textcolor{black}{Hamming Distance Based Approach}}
In \cite{IEEEexample:daugman2003}, Daugman proposed a method for measuring uniqueness of iris texture and evaluated it on a large collection of iris scans.
The method is based on a feature representation in which each iris scan is transformed into a 2,048-dimensional binary vector (referred to as an iris code) by applying a multi-scale wavelet decomposition to iris textures and encoding the resulting phase characteristics.
\textcolor{black}{
The gist of the estimation method relies on \textcolor{black}{the} comparison of pairwise distances measured between actual feature vectors and synthetically generated ones.
The elements of synthetic vectors are drawn from a binary distribution in an \textcolor{black}{independent and identically distributed} (iid) manner to establish a basis for comparison.
}
By interpreting the match between each element of two subject's feature vectors as a Bernoulli trial, the total number of matches is expressed as a random variable. 
This effectively corresponds to the Hamming distance between the two vectors which is known to follow binomial distribution under the iid assumption.

The binomial distribution can be characterized by the number of elements $N$, the probability of success in each trial $p$, and the variance of the number of matches $\sigma^2$ as
\begin{equation}
\label{eq:daugman}
N = p(1-p)/{\sigma^2}.
\end{equation}
Hence, for an empirically obtained distribution, measured mean and variance values (which are estimators for $p$ and $\sigma^2$) can be used to determine the number of iid elements, $N$, in the vector.
Using this formulation, Daugman computed normalized Hamming distances between iris codes of $4,258$ subjects in a \textcolor{black}{pairwise} manner which yielded more than 9 million comparisons.  
Then, the resulting mean and variance values are evaluated to determine the corresponding degree of freedom in a binomial distribution, {\em i.e.} equivalent $N$ that will yield the same statistic from iid binary vectors, which is found to be 249.
Hence, the binary iris code is estimated to have 249 bits of entropy. 

This method of estimating uniqueness has certain limitations. 
The reliability of estimation depends on the underlying dependency of the feature vector array. 
Our observations show that for apparent forms of dependencies, such as repetition of the elements in a feature vector, this approach is effective. 
However for more subtle dependencies, say, XOR'ing the first half of the vector with the second half and appending it to the vector, the measured degree of freedom increases proportionally. 
Therefore, the method has a tendency to overestimate the number of independent elements in the vector.

This approach also assumes that each element of the feature vector is equally important as the contribution of each feature to \textcolor{black}{the} Hamming distance is equally weighted. 
Hence, it cannot be generalized to representations where features are sorted \textcolor{black}{according to} their importance. 
In addition, this type of modeling holds only when each element of the feature vector is identically and uniformly distributed. 
Otherwise, the relation given in Eq. (\ref{eq:daugman}) does not hold.  
But most critically, since this approach relies on evaluating \textcolor{black}{pairwise} differences between feature vectors, it cannot incorporate within-subject 
variations into its formulation. 
In this sense, it is more suitable for biometric modalities where within-subject variability is very limited.

\vspace{-0.3cm}
\subsection{Measuring Relative Entropy in the Feature Space}
An alternative biometric information measurement approach is proposed by Adler {\em et al.} \cite{adler2009towards, IEEEexample:adler2006}  considering facial images and using a relative entropy based formulation.
Relative entropy (also known as Kullback-Leibler divergence) is a non-symmetric measure of the \textcolor{black}{distance between two probability distributions, and it
estimates the average number} of additional bits required to code samples from one distribution when using a code optimized for the other distribution.
This approach effectively \textcolor{black}{defines} biometric information as \textcolor{black}{the additional amount of information needed to describe each individual when 
a collection of biometric features that represent population characteristics are known. 
This is expressed in terms of the relative entropy formulation as}
\begin{equation}
D(p||q) = \int_\mathbf{X} p(\mathbf{X}) \log_2 \frac{p(\mathbf{X})}{q(\mathbf{X})} d\mathbf{X}
\label{eq:RE}
\end{equation}
\textcolor{black}{where $\mathbf{X}$ is a multi-dimensional feature representation, $q(\mathbf{X})$ is the distribution of $\mathbf{X}$ in the overall population, and $p(\mathbf{X})$ denotes the distribution of a subject's features.
By assuming multivariate normality for distributions of eigenface features, 
average relative entropy between the population distribution and the individual distributions is computed as an upper bound to biometric information content of face images.
Later, \cite{IEEEexample:entropy2015} applied this approach to voice biometrics considering i-vector feature representation of voice.}

One limitation of this formulation concerns the fact that modeling within-subject variability requires estimation of too many parameters ({\em e.g.}, mean vector and covariance matrix considering a multivariate normal distribution) which becomes highly error prone when there are only a few samples from each subject.
\textcolor{black}{Therefore, this approach requires deployment of regularization schemes to guard against numerical instability \cite{adler2009towards, IEEEexample:entropy2015}.}
Further, when actual distributions are not normally distributed, the estimation becomes less reliable.  
Aside from modeling and computation difficulties, this definition of biometric information has a tendency to overestimate the \textcolor{black}{level of distinctiveness} as relative entropy provides a measure depending on how a subject is different from the population, yet, it does not capture the fact that two subjects can sufficiently be different from the population but might still be very alike.

\vspace{-0.2cm}
\subsection{Measurements in the Matching Score Space}

Another estimation approach aims at measuring the average information provided by a speaker verification system utilizing matching scores computed by the system.  
In \cite{ramos2008cross}, authors investigated the amount of information needed by a speaker verification system to make a decision as to whether two utterances match. 
Accordingly, the information gain provided by the system is estimated in terms of the cross-entropy between empirical distributions of scores for matching and non-matching speakers and 
the distribution of a variable indicating the prior belief in correctness of a match decision.

Later, Takahashi {\em et al.} \cite{IEEEexample:takahashi2010} and Sutcu {\em et al.}\cite{sutcu2010measure} independently introduced an estimation approach to alleviate the limitation of the relative entropy-based measurement approach due to lack of \textcolor{black}{a} sufficient number of subject samples needed for modeling.
\textcolor{black}{To this end, 
\cite{sutcu2010measure} proposed computing the relative entropy between within-subject and between-subject distributions of scores generated by the biometric matching system, rather than using feature distributions directly.}
Alternatively, \cite{IEEEexample:takahashi2010} provided a mutual information based formulation that asymptotically approximates relative entropy in order to obtain the entropy of the biometric system.

Since these measures \textcolor{black}{only rely on} score distributions, they effectively reduce high-dimensional feature distributions given in Eq. (\ref{eq:RE}) to single dimensional score distributions.
Hence, they do not suffer from difficulties of estimating parameters of multivariate distributions.
In addition, within-subject variability can be captured more accurately as the number of data points needed for modeling increases in proportion to the square of the available number of samples due to pairwise comparison of subject samples. 
Overall, this resulted in a more tractable measure where the reliability of estimates mainly \textcolor{black}{depends} on accuracy of the two distributions.
The obvious shortcoming of this type of approach is that it is not a true measure of biometric information content as it also depends on the matching algorithm used for evaluating similarity or closeness between biometric features. 
However, achievable distinguishability within the confines of a biometric identification system is also a key consideration in practical settings.

\vspace{-0.3cm}
\subsection{Applicability of Existing Measures to Voice Biometrics}

To determine the uniqueness of a feature representation of speech, limitations and strengths of these measurement approaches must be evaluated in the context of established feature representation for voice.   
In this regard, an important attribute of i-vectors is that their elements can be assumed independent
because the total variability matrix involved in their calculation can be regarded as an eigenspace with i-vectors functioning as eigenvectors 
\textcolor{black}{and further because i-vector elements are decorrelated using a whitening transformation and length normalization \cite{garcia2011analysis}.}
Another attribute concerns the fact that i-vector elements are sorted based on their ability to distinguish speakers as they are obtained through linear discriminant analysis. 
As a result of this, error rates of speaker verification systems do not decrease linearly with decreasing dimensionality of i-vectors.
Lastly, elements of i-vectors are continuous valued and are modeled as a Gaussian mixture distribution as part of the GPLDA based matching process. 

Considering the overall characteristics of i-vector representation, 
these measures have some shortcomings with respect to their applicability to assess their uniqueness.
In the case of Hamming distance based approach \cite{IEEEexample:daugman2003}, not only each i-vector element has a different discriminative power
but also their quantization will not yield the required uniform distribution.
Further, i-vectors have relatively high within-speaker variability which cannot be adequately captured by this approach. 
Although the relative entropy based estimation approach \cite{IEEEexample:adler2006} \cite{adler2009towards} does not require quantization of i-vector elements, its formulation is more sensitive to modeling errors \textcolor{black}{as distribution of feature vectors at a high-resolution representation cannot be empirically obtained.}
Since computation of Eq. (\ref{eq:RE}) involves division of two distributions, calculations are more prone to errors at distribution tails where values are small and accurate modeling is typically challenging due to the limited number of samples per speaker.
Moreover, i-vector elements follow a Gaussian mixture model which further increases the number of parameters to be correctly determined.
\textcolor{black}{Finally, the measurement approach utilizing distance or score distributions between biometric features, \cite{IEEEexample:takahashi2010, sutcu2010measure}, crucially estimates the distinguishability with respect to a similarity or distance metric.} 
\textcolor{black}{This approach potentially underestimates the distinguishability intrinsic to the feature representation as it essentially operates on a function of feature vectors.}

Inspired by the mutual information based formulation of Takahashi {\em et al.} \cite{IEEEexample:takahashi2010}, we introduced a new approach for estimating biometric information content of human voice using i-vector representation \cite{tandogan2017towards}.
\textcolor{black}{Most notably, in this approach, each i-vector is viewed as an instance of a discrete multivariate random variable to obtain a more tangible uniqueness measure defined in terms of the number of bits required to represent i-vectors.
By discretizing feature vectors, this approach allows for more accurate estimation of feature distributions instead of assuming a particular distribution model for features. 
Further, since measurements are performed on actual feature vectors, as opposed to utilizing scores or distances computed between features, this measure provides a more accurate reflection of biometric information content. }
This study further expands on this initial work to investigate the effects of i-vector discretization more systematically and 
to determine how factors critical to modeling of between-speaker and within-speaker variability affect uniqueness estimates.
The details of our approach are discussed in the following section.

\vspace{-0.2cm}
\section{Measuring Mutual Information in Feature Space}
\label{sec:method}
Biometric information quantifies the ability to uniquely identify subjects through their biometric traits. 
In the case of voice biometrics, this reduces to the uncertainty in the composition of i-vectors while disregarding speaker level variations. 
This definition can indeed be related to the concept of entropy.
\textcolor{black}{Essentially, entropy measures the amount of uncertainty associated with random variables and shows the average number of bits needed to represent each possible
outcome of a discrete variable.}
\textcolor{black}{For a random variable $X$ that takes $k$ different values with probabilities $p_1, p_2, \ldots, p_k$, the entropy is defined as
\begin{equation}
    H(X)=\sum_{i=1}^{k} p_i \times \log_2 \frac{1}{p_i}.
\label{eq:entropy}
\end{equation}
However, the entropy of a feature representation for a modality by itself does not correspond to the biometric information content as it does not capture the within-subject variability.}

Given a random variable $S$ representing a randomly selected speaker among a group of $n$ speakers $\{s_1,\ldots,s_n\}$ and an \textcolor{black}{m-dimensional} multivariate random variable ${\bf V}=[V_1,\ldots,V_{m}]$ whose realizations represent, say, i-vectors of speakers, the degree of dependence between the two variables 
provides a measure of intrinsic distinguishability associated with i-vector representation. 
In fact, the more interrelated the identity of a speaker to his/her i-vectors, the higher is the biometric information content provided by the representation. 
Similarly, if the two are less dependent, the biometric representation will be less discriminative of the speaker identity and, thereby, will yield lesser overall information. 
Hence, the mutual information between $S$ and ${\bf V}$, can be used to evaluate the biometric information content similar to initially formulated in \cite{IEEEexample:takahashi2010} as,
\begin{equation}
    I(S;{\bf V})=H(S)-H(S|{\bf V}),
\label{eq:mut}
\end{equation}
where $H(S)$ denotes the entropy of $S$ and $H(S|{\bf V})$ is the corresponding conditional entropy expressing the average uncertainty in speaker identities given
the population characteristics. 
Takahashi {\em et al.} \cite{IEEEexample:takahashi2010} argued that this quantity asymptotically approximates the relative entropy and used within-speaker and between-speaker distributions of matching scores to evaluate it.

This formulation effectively measures the decrease in uncertainty about the identity of speakers due to known aggregate characteristics.
However, there are a number of challenges in evaluating Eq. (\ref{eq:mut}). 
First, the probability distribution describing the uncertainty of speaker identities is not known.
In the absence of this distribution, all speakers can be assumed to be equally likely to be identified, thereby, maximizing $H(S)$ and potentially leading to an overestimation in calculations. 
Second, the evaluation of conditional entropy, $H(S|{\bf V})$, crucially requires recomputing the distribution of uncertainties concerning speaker identities for a given i-vector, as speakers with i-vectors distributed in that locality of the i-vector space will be better identifiable.

To avoid these complications, in our approach, we utilize the alternative derivation for $I(S;{\bf V})$\textcolor{black}{\cite{cover1999elements}}, expressed as 
\begin{equation}
    I(S;{\bf V})=H({\bf V})-H({\mathbf V}|S),
\label{eq:mut2}
\end{equation}
where $H({\bf V})$ corresponds to the entropy of i-vectors and $H({\bf V}|S)$ 
is the corresponding entropy conditioned on speaker identity, {\em i.e.} average entropy of each speaker's i-vectors.
\textcolor{black}{The two terms in Eq. (\ref{eq:mut2}) essentially relate to between-speaker and within speaker variations.
In this regard, $H({\bf V})$ can be interpreted to measure between-speaker variability as it is computed over all speaker i-vectors. 
By itself, however, this quantity overestimates the entropy of the i-vector feature representation as it implicitly assumes speaker i-vectors are invariant.
In contrast, $H({\bf V}|S)$ captures within-user variability and offsets for this error by eliminating the contribution of speaker level variations.
}
Both of these quantities, at their core, require estimating feature distributions. 

An important consideration here is that the above formulation does not yield a tangible information measure when applied to continuous variables, such as i-vectors whose elements take real \textcolor{black}{values}.  
\textcolor{black}{In contrast, for discrete random variables, entropy indicates the lower bound for the expected number of bits required to express each instance of variable ${\mathbf V}$ or ${\bf V}|S$.
This interpretation is central to our definition of biometric information content.}
Therefore, unlike the continuous representations used in speaker verification systems, i-vectors need to be discretized. 
\textcolor{black}{
This in turn requires determining the degree of discretization that must be applied to i-vectors while still preserving their intrinsic distinguishability.
Ideally a very high-resolution representation is preferable, but when feature distributions have to be estimated empirically this will require a large amount of biometric data.
Since this is hard to attain, one can alternatively decide on the right level of quantization by comparatively evaluating the speaker verification performance 
obtained under different quantization settings. 
}

An i-vector can be discretized using element-wise scalar quantization as its elements are decorrelated.
Moreover, the distribution of i-vector elements are highly non-uniform; therefore, minimizing the error ({\em i.e.,} information loss) due to quantization is critical to retain an accurate representation.
This can be realized by using optimal quantization methods, such as Lloyd-Max quantizer \cite{max1960quantizing}, that can better adapt to the distribution of the i-vector elements.

Ultimately, to evaluate Eq. (\ref{eq:mut2}) both $H({\bf V})$ and $H({\bf V}|S)$ must be computed. 
\textcolor{black}{For the general case, assuming ${\bf V}$ is composed of $m$ independent components, {\em i.e.,} $[V_1,V_2,...,V_{m}]$, $H({\bf V})$ can be calculated as the sum of the entropy of each i-vector element as}
\begin{equation}
H({\bf V})=\sum\limits_{j=1} ^{m}H(V_j)=H(V_1)+\ldots+H(V_{m}).
\label{eq:ivector_entropy}
\end{equation}
Similarly, the conditional entropy $H({\bf V}|S)$ can be calculated by taking an average over all speakers in the dataset as
\begin{equation}
H({\bf V}|S)=\sum\limits_{i=1}^n P(s_i)H({\bf V}|S=s_i)
\label{eq:average}
\end{equation}
where $P(s_i)$ is the probability of encountering speaker $s_i$ among $n$ speakers wherein each speaker can be considered equally likely, {\em i.e.}, $P(s_i)=\frac{1}{n}$,
and $H({\bf V}|S=s_i)$ is the entropy in speaker $s_i$'s i-vectors.
Hence, by substituting Eqs. (\ref{eq:ivector_entropy}) and (\ref{eq:average}) in Eq. (\ref{eq:mut2}), 
the distinctiveness provided by the i-vector representation can be finally calculated as
\begin{equation}
I(S;{\bf V})=\sum\limits_{j=1}^{m} H(V_j)-\frac{1}{n}\sum\limits_{i=1}^n\sum\limits_{j=1} ^{m} H(V_j|S=s_i). 
\label{eq:final}
\end{equation}

\textcolor{black}{Obviously, evaluation of this expression requires i-vector distributions for the overall population, {\em i.e.}, $p(\mathbf{V})$, and individual speakers,  
{\em i.e.}, $p(\mathbf{V}|s_i)$.
In the absence of parametric models, underlying i-vector distributions must be obtained empirically.
This can be performed reliably only if both a large number of speakers and a large number of speech samples per speaker are used for modeling}.
We next describe the datasets used in our measurements along with details concerning their creation.

\vspace{-0.2cm}
\section{Datasets}
\label{sec:dataset}

Reliable estimation of uniqueness ultimately comes down to whether the data used for modeling speakers accurately capture between- and within-speaker variations.
Although several speech corpora have been used for benchmarking the performance of speaker verification methods, the number of speakers and the duration of speech samples from each speaker included in most of these datasets do not provide adequate data points needed for accurate evaluation of Eq. (\ref{eq:final}). 
Moreover, speech utterances included in these datasets do not sufficiently exhibit the natural variation present in a speaker's voice 
as they are captured under well defined settings. 

To better address these limitations, in this work, we use the public VoxCeleb dataset \cite{Nagrani17} \cite{Chung18b} and introduce two new datasets that include speech samples collected from publicly available sources.
Although a speech corpus drawn entirely from public data sources does not provide a control over how samples are collected, it allows the creation of a large-scale and diverse dataset as needed by our formulation. 
The following subsections provide details on how these datasets are generated\footnote{Both of the newly created datasets will be made publicly available following final modifications of this manuscript.}.

\vspace{-0.3cm}
\subsection{TEDx Dataset}
In our earlier work \cite{tandogan2017towards}, we performed measurements on a corpus obtained from TED Talks which involves a library of videos in which speakers deliver monologue-style presentations on a wide variety of topics.
The online archive for TED Talks provides rich metadata about the talks and speakers, the audio captions, and the option to choose among a variety of high quality audio recordings.
However, the available number of videos is limited to only a few thousand. 
To create a more diverse dataset, in this work, we utilized TEDx Talks which follows a similar format and the same rules as the original TED Talks.  
Since TEDx events are organized independently, its archive involves a larger collection of talks in a variety of languages \cite{tedxTalks}.

TEDx videos are featured on the TEDx channel of the YouTube video sharing website.
The durations of these videos range from a few minutes to up to an hour, with most talks lasting around 20 minutes. 
Although it is not possible to ascertain recording conditions for these talks, a great majority have an audio bitrate around 120 Kbps, 
which is most likely due to YouTube re-encoding of all uploaded videos. 
To create our dataset comprising speech samples of TEDx speakers, we first obtained URLs of all videos by going through all available TEDx playlists.
We then examined the index page of each video by searching video metadata for the content tag in order to identify the talks in English language.
We disregarded all videos that lack a caption file and identified 24,500 videos whose audio tracks and subtitles were later downloaded using the youtube-ld download tool \cite{youtubeDl}.

The TEDx presentations are given by a single speaker, meaning speech overlap from multiple speakers is not a concern. 
However, identification and extraction of speech-only segments from the obtained audio tracks requires further effort.
To this end, we utilized the audio captions associated with each video in conjunction with a text-to-speech alignment tool.
By using the CMU Sphinx Aligner Toolbox \cite{lamere2003cmu}, we identified time intervals when each word in the caption is spoken in the source audio. 
Essentially, this enabled us to remove all non-speech utterances like music, applause and silence from audio samples, as well as all speech utterances where background noise masked their audibility.

With this tool, the total processing time spent during alignment of each audio track is found to be proportional to the overall length of the audio track.
However, the tool would occasionally take too long to produce a result or would fail to return results due to an error.
Therefore, we eliminated all tracks whose processing took more than twice the duration of the input audio track, leaving us with 22,598 tracks. 
Then, all speech utterances aligned with caption words are subjected to voice activity detection \cite{sohn1999statistical} to 
eliminate silence intervals between utterances. 
The resulting speech segments are then combined together to obtain a speech-only audio track from each video. 
From the resulting tracks, we eliminated all those that yielded less than 5 seconds of speech utterance, and the remaining ones are partitioned into samples with lengths around 5 seconds.
This overall led to a collection of 1,625,915 speech samples associated with 20,741 speakers with varying numbers of samples per speaker.

\vspace{-0.8cm}
\textcolor{black}{
\subsection{VoxCeleb Dataset}
This dataset consists of over a million utterances from over seven thousand speakers.
The dataset was released in two stages, referred to as VoxCeleb1 \cite{Nagrani17} and VoxCeleb2 \cite{Chung18b}, which respectively included 1,251 and 6,112 speakers. 
The utterances are extracted from YouTube videos that include interviews, excerpts, and public speeches shot at a wide variety of events and circumstances. 
When obtaining utterances, speakers in a video are tracked by detecting and verifying their faces.  
Then, each utterance is associated with a speaker based on the synchronization of the mouth-motion and speech in the video to eliminate segments that contain dubbing or voice-over. 
The durations of utterance vary from three to nine seconds with an overall average close to five seconds.
}

\textcolor{black}{
In regards to their collection, VoxCeleb and TEDx datasets are very similar. 
However, the speech utterances in the TEDx dataset are expected to be more accurately identified as they are extracted from single speaker videos based on successful speech-to-text alignment as opposed to using mixed type of videos while relying on audio-visual synchronization.
In contrast, since the VoxCeleb dataset includes speech samples recorded over multiple sessions,   
it is expected to exhibit more diverse characteristics than the TEDx dataset which essentially includes a single session per speaker.}
For our experiments, all utterances are subjected to voice activity detection to remove silent parts, and the very short ones (that won't allow extraction of a single i-vector) are eliminated. 
Measurements are performed on all of the remaining speech samples.

\vspace{-0.3cm}
\subsection{Movie Dialogues Dataset}

Similar to various other datasets, the TEDx dataset involves speech samples expressed in a limited emotional tone of voice ({\em i.e.}, dominated by presentation voice) and does not incorporate the emotional range intrinsic to a speaker's voice. 
Hence, an attempt to estimate the individuality of the human voice solely using such a corpus will undoubtedly result in overestimating \textcolor{black}{the degree of individuality regardless of which biometric feature representation is used}.  
To partially address this challenge, we created another dataset comprising speech samples extracted from movies. 
Since movies are typically composed of dialogues between two or more speakers carried out under various environmental circumstances, they provide a better basis for capturing
the within-speaker variability.

Therefore a similar approach based on alignment of audio with the movie caption is deployed to obtain speech samples.
The main challenge here, however, concerns correct attribution of each speech utterance to its speaker. 
Although movie subtitles follow some style rules, they don't necessarily identify speakers individually in the text-dialog.
Most generally markers, such as hyphens, are used to denote dialogue without including speakers' names.
Even when identifiers are used, they may be excluded if the speaker is visually apparent in the corresponding time-synchronized video frames.
Furthermore, descriptions for non-verbal sounds may also be included as part of subtitling.  
Therefore, before an alignment is performed, the speech segments associated with each speaker must be determined. 
One way to realize this is through clustering of utterances based on a speaker verification approach. 
That is, creating a model for each speaker and then verifying the source of each utterance.
However, automatic creation of speaker models is error prone as it must be done incrementally, especially for actors with fewer lines of dialogue. 
Therefore, we considered utilizing movie scripts, which are written versions of what happens in a movie, in conjunction with movie subtitles to correctly attribute each part of a dialogue to a speaker.

For this purpose, we determined public sources on the Web that archive movie scripts and screenplays\footnote{We identified following websites as potential sources for movie scripts with the first one identified to provide the most comprehensive collection.\\
https://imsdb.com \\
http://www.dailyscript.com/movie.html\\
http://www.simplyscripts.com/movie-screenplays.html\\
http://www.awesomefilm.com}.
Examining these collections, we identified more than a thousand movies from which to extract dialogue samples.
We then retrieved these scripts and manually eliminated those that were in scanned document format 
and those that do not explicitly designate the speaker for each speech segment.
In addition, movie soundtracks are extracted from their DVD formatted versions using the FFMPEG video processing tool along with their subtitles. 
The retrieved scripts are then checked against actual dialogues of soundtracks for potential discrepancies.
The comparison of subtitles and the movie scripts of several movies revealed further differences both at the narrative-level, due to missing or extra lines, and at the sentence-level, where similar meaning was conveyed with a different sentence construction or choice of words.
We determined that these differences were essentially due to scripts being draft versions and not the final shooting scripts.

All styles of subtitling utilize line-breaks to segment speech, and when multiple speakers are present in a scene they are separated by breaks.
Therefore when attributing speech segments, each text-line or full-sentence (when punctuation is used) in the subtitles are used as the basis of search. 
Since smaller phrases are likely to yield various matches in the script, we initially identified all uniquely matching lines and sentences in the script along with their speakers.
Then, treating those exact matches as reference points, subsequent text segments in the script are searched only within a limited range in subtitles, thereby restricting probability of false attributions.
Each of the remaining lines in the movie subtitle is then attributed to a speaker by evaluating its similarity to text in the script \cite{javaStringSimilarity}. 

When computing the similarity of two strings, each text-line in subtitle is matched against text-strings that may be shorter or longer by two words. 
For this purpose, we first performed a string comparison using the Levenshtein distance measure. 
We empirically determined similarity thresholds of \textcolor{black}{85\%} or above in order to accept a match and 40\% or below to eliminate a line from matching. 
Remaining unattributed lines are subjected to further comparison.
To overcome potential spelling errors, we first utilized the Jaro-Winkler distance measure to compare words in the subtitles and the script, and two words with comparison values of more than 95\% are considered to be the same.  
Among the remaining lines, those for which string search yielded a Jaccard similarity above 50\% are considered matched and attributed to the corresponding speaker.  

After attributing speech segments in the subtitle to speakers denoted in the movie script, we used the Sphinx tool to align text with audio just as before to identify each utterance corresponding to spoken words in the subtitle. 
Finally, we utilized the IMDB cast lists to identify actors corresponding to speakers in each movie and to consolidate speech samples of actors obtained from different movies. 
Extracted speech utterances are then partitioned into 5-second long samples after voice activity detection.
At the end of this overall process, we were able to obtain 21,523 speech samples of 1,595 actors from 249 movies where 556 actors had at least 10 samples, 286 actors had more than 20 samples and 132 actors had more than 40 samples.

We must note here that the Movie Dialogues dataset resulted in fewer speakers than expected due to two main factors.
First is due to the inability to access final versions of movie scripts which would have matched exactly with the movie subtitles and enabled us to attribute each utterance to its speaker. 
In their lack, to prevent false-attributions as much as possible, our association method was essentially tuned to eliminate text-lines if there is ambiguity when evaluating similarity within draft scripts.
The second factor is due to performance of the aligner which performed considerably worse on movies as compared to TEDx videos due to higher interference from background noise, sound effects, and simultaneous dialogues. 
Nevertheless, this dataset is unique in its composition and its attempt to capture true within-speaker variability in the human voice.

\vspace{-0.2cm}
\section{Evaluation of the Uniqueness Measure}
\label{sec:results}
To evaluate our measure, we first train a GMM-UBM based i-vector system using the MSR Identity Toolbox \cite{IEEEexample:msr2013}.
For this, we compute MFCC features from all speech samples using a 25 milliseconds sliding Hamming window at intervals of 10 milliseconds.
The 19 MFFCs extracted from each window are further expanded with the log energy of the window as well as delta and acceleration coefficients (first and second order derivatives computed over time).
The resulting 60 dimensional MFCC features are then used to develop a 512-component GMM-UBM model to extract speaker i-vectors as described below.

\vspace{-0.3cm}
\subsection{Evaluation Protocol}

\textcolor{black}{
For our experiments, the datasets are split into development and measurement sets with disjoint speakers. 
The speech samples in the development set are used for training the UBM, total variability matrix, LDA and GPLDA.  
Using these models, the speaker samples in the measurement set are transformed to the i-vector feature space to obtain uniqueness estimates.
The measurement set is also used to benchmark the identification performance at different levels of quantization of i-vectors.
Therefore, it is further divided into two as enrollment and test subsets. 
For identification, the training (enrollment) and the tests are performed on the same group of speakers.
}

\textcolor{black}{
The development partition of the TEDx dataset includes 48,133 speech samples from 5,000 speakers with an average of 9.6 samples per speaker. 
Correspondingly, the measurement set includes the remaining 15,741 speakers and the associated 1,577,782 samples with around 100 speech samples per speaker.
In the case of VoxCeleb dataset, a subset of the VoxCeleb1 dataset that includes 10 samples from each of the 1,251 speakers is designated as the development set, 
and the $784,312$ samples associated with the 6,060 speakers in the VoxCeleb2 dataset are used as the measurement set.
As for the Movie Dialogues dataset, due to its relatively small size which includes 21,523 samples from 1,595 speakers, it is solely used as a measurement set, 
and the development parameters obtained for the TEDx dataset are used to extract speaker i-vectors.   
}

\textcolor{black}{
For the two development sets, UBMs are separately generated using the available speaker samples. 
Then, the total variability matrices are estimated through five iterations of the expectation-maximization (EM) algorithm.
Using the resulting total variability matrices, i-vectors for all speech samples in the two development sets are computed. 
These i-vectors are then used to estimate the LDA matrices. 
The i-vectors projected with LDA down to 200 dimensions are then centered, length normalized and whitened.
Finally, the post-processed i-vectors are used to estimate the GPLDA model by 10 iterations of the EM algorithm. 
}

\textcolor{black}{
When assessing verification performance all but one of the speech samples of all the speakers are used to create an i-vector-based speaker model, and the i-vector obtained from the remaining test sample is used for testing. 
Each speaker model is created using two approaches \cite{IEEEexample:msr2013}. 
In the first one, i-vectors extracted from all samples are averaged together to obtain one speaker i-vector.
In the second one, MFCC features are averaged across multiple samples of a speaker and an i-vector is extracted from these averaged MFCC features. 
When computing error rates the test i-vector is compared against each speaker model in the test set using the GPLDA model.  
The resulting matrix of decision scores is then used to determine the thresholds needed for verifying the speaker of a test i-vector and to compute the false-acceptance and false-rejection rates in speaker verification.
}

\vspace{-0.3cm}
\subsection{Quantization of I-Vectors}
Our uniqueness measure is defined in a discrete feature space; therefore, measurements are performed on quantized versions of i-vectors in the measurement set.
Since quantization operation incurs information loss, it is expected to cause a decrease in the individuality inherent to i-vectors.
\textcolor{black}{However, at the same time, it is very plausible that the i-vector based speaker verification and identification systems do not fully exploit the information in the  high resolution representation of i-vectors and quantization will not cause a loss in accuracy.} 
Therefore, it is important to determine the right level of quantization that can be performed while ensuring a comparable performance when operating on original and quantized 
i-vectors.

\textcolor{black}{To assess the impact of quantizing i-vectors, we utilize the speaker verification performance as the basis of evaluation.}
\textcolor{black}{That is, by applying different levels of quantization to i-vectors, we determine how verification performance changes with respect to the use of original i-vectors. 
As the first step for this, the pre-computed i-vectors in the development set are examined to learn their distribution and determine the best suited quantization scheme. }
Since i-vector elements are uncorrelated there is less to be gained from vector quantization.
In addition, the distribution of i-vector elements is highly non-uniform; therefore, we utilize the Lloyd-Max algorithm \cite{max1960quantizing} to create an optimal partitioning (in the mean-squared error sense) of the i-vector elements for a given number of quantization levels ({\em i.e.,} bits per quantized sample). 
\textcolor{black}{These i-vectors are then quantized using the learned parameters for each quantization setting, and the LDA matrix and GPLDA model parameters are re-estimated.}
Using the same quantization parameters, the i-vector speaker models and the test i-vectors are also discretized, and the error rates are computed in the same manner.

We use the equal error rate (EER) as the performance metric for speaker verification, which refers to the point where false-acceptance and false-rejection rates are equal. 
The EER values obtained under different quantization settings for TEDx and VoxCeleb datasets are given in Table \ref{tab:quantizationEER}. 
In the table, the third column shows EER values \textcolor{black}{obtained} when original, non-quantized i-vectors \textcolor{black}{(expressed by a 32-bit floating-point representation)} are used and subsequent columns correspond to increasing numbers of quantization bits. 
The two error values, $EER_1$ and $EER_2$, respectively correspond to cases where i-vectors and MFCC features are averaged when creating speaker models. 
As the corresponding values indicate, when i-vector values are quantized at low resolution ({\em i.e.,} 1 or 2 bits), EER values are higher than the 
high-resolution float representation.
This is expected as severe quantization suppresses both between-speaker and within-speaker variability in i-vectors, making them less distinguishable.  

For higher quantization levels, however, we observed an interesting phenomenon where quantization of i-vectors yielded a slightly improved EER than the original i-vectors.
This essentially indicates that at 2-5 bits non-uniform quantization of i-vector elements, the gain obtained due to decrease in within-speaker variability compensates for the errors due to decrease in between-speaker variability. 
In other words, quantization enabled slightly better clustering of speaker i-vectors while still preserving relative distances between different speakers. 
We also determined that the use of n-bit codewords to represent quantized i-vector elements (where n is the number of quantization bits), instead of using actual quanta values associated with each quantization interval, will also induce a similar effect.
Our comparison of the two cases on the TEDx dataset reveals that codeword-based representation yields lower within-speaker variance in the GPLDA model which in turn causes around 6\% decrease in measured EER values.

We must note that this observation is in agreement with results of our earlier work \cite{tandogan2017towards} as given in the last row of the table. 
This work utilizes a collection of speech samples obtained from TED Talk videos of 1,914 speakers by following the same procedure described for creation of the TEDx dataset. 
In comparison, these videos feature higher quality audio recordings with very accurate transcripts. 
To measure the EER values, speech samples from 993 speakers are used as development data and the remaining samples from 921 speakers are used for measurements and tests.

As expected, at higher quantization bit levels, EER values will eventually approach the non-quantized case. 
This trend is observed for the TEDx dataset; however, for VoxCeleb and TED datasets there are slight fluctuations in the measurements.
This may be attributed to the limited number of speakers used for development and tests, which makes EER computations less accurate.
Overall, based on these results, we deduce that distinctiveness provided by the i-vector feature representation 
can be evaluated considering 2-5 bit quantization of i-vector elements, with 3-bit quantization yielding more reliable estimates, as these quantized representations yield comparable or marginally better EER values in speaker verification as compared to using original i-vectors.

\def\arraystretch{1.1}
\begin{table}[!ht]
	\centering
	\caption{Change In Verification Performance Due To I-Vector Discretization for Two Speaker Model Generation Approaches
	($EER_1$: Averaging i-vectors, $EER_2$: Averaging MFCC coeffs.)}
	\begin{tabular}{cccccccc}\cline{3-8}
		
        \hline
		 \multicolumn{2}{r}{\textit{\# of bits:}} & \textit{$float$} & \textit{1} & \textit{2} & \textit{3} & \textit{4} & \textit{5} \\

		\hline
		\hline
		\multirow{2}{3em}{TEDx} & {\textit{$EER_1$}}	& 2.73 & 4.77 & \textbf { 2.46 } & 2.56 & 2.66 & 2.81\\
		 & {\textit{$EER_2$}}	& 2.39 & 4.96 & 2.56 & \textbf {2.36} & 2.38 & 2.45\\
		\hline
		  \multirow{2}{4em}{VoxCeleb} & {\textit{$EER_1$}}	& 8.03 & 8.33 & 6.15 & \textbf { 6.10 } & 6.32 & 6.28\\
		   & {\textit{$EER_2$}}	& 7.25 & 8.62 & 6.12 & 5.94 & 6.09 & \textbf{ 5.91 }\\
		  \hline 
		  TED  \cite{tandogan2017towards} & {\textit{$EER$}} &	2.89 & 4.77 & 2.38 & \textbf {2.06} & 2.17 & 2.17\\
		  \hline
	\end{tabular}
	\label{tab:quantizationEER}
\end{table}

\vspace{-0.3cm}
\subsection{Uniqueness Estimates}
After discretization of i-vectors into a fixed number of quantization levels, we estimate the uniqueness in human voice using i-vector representation as described in Sec. \ref{sec:method}.
\textcolor{black}{This is realized by first empirically obtaining the distribution of i-vector elements and then
computing the entropy of overall i-vectors, $H({\bf V})$, and the entropy of a speaker's i-vectors, $H({\bf V}|S)$, 
to finally obtain the biometric information content of i-vector representation, $I(S;{\bf V})$, as defined in Eqs. (\ref{eq:mut2})-(\ref{eq:final}).}
Calculated uniqueness estimates on TEDx and VoxCeleb measurement sets are provided, respectively, in Tables \ref{tab:samples-p-speaker} and \ref{tab:VoxCeleb2}.  
The first two rows of both tables show the number of samples per speaker that estimates are based on and the available number of speakers with that many samples.
Subsequent rows give estimates for increasing number of bits used to represent each quantized i-vector element.
Similarly, columns present uniqueness estimates under increasing numbers of samples from speakers 
to reflect the effect of better capturing the within-speaker variability.

As can be seen in these tables, estimates obtained from the two datasets are similar and exhibit the same trends for increasing numbers of quantization bits and samples. 
In both tables, it is observed that estimates start from 20-40 bits for 2-bit quantization and gradually increase to several hundred bits for the 5-bit quantization setting.
This increase in the uniqueness estimates, however, does not translate into better discrimination of speakers.
As demonstrated in the results of Table \ref{tab:quantizationEER}, in terms of the achievable EER performance, 2-5 bit quantization yields the same level of distinguishability as using the original i-vectors.
That is, using a higher resolution representation of an i-vector does not lead to a discernible improvement in verification but rather contributes to randomness in i-vectors. 
It must also be noted that estimates become more error-prone with increasing numbers of bits because the same number of samples are used to obtain feature distributions for higher resolution representations.
In the case of the number of samples per speaker, we see that estimates converge closely when a large number of samples are used for estimation.
More importantly, it is seen that estimates based on 10 samples are roughly 1.5-4 times higher than those obtained using 90-100 samples, showing how fewer samples per speaker may lead to a significant overestimation of the biometric information content.

Overall, our measurements estimate the uniqueness of the i-vector representation to be in the range of 42-75 bits when the TEDx dataset is taken as basis and 37-70 bits for the VoxCeleb dataset, depending on the number of bits used to represent each i-vector element. 
The difference in estimates between the two datasets can mainly be explained by the fact that speech samples in the TEDx dataset are obtained under a more controlled environment with lesser within-speaker variability, thereby yielding higher estimates.
Next, we examine in more detail how different factors, such as number of speakers, sample duration, and the extent of within-speaker variability, affect uniqueness estimates. 

\begin{table*}[!ht]
    \scriptsize
	\centering
	\caption{Uniqueness Estimates from TEDx Dataset for Increasing Number of Samples}
	\label{tab:samples-p-speaker}
	\begin{tabular}{c|cccccccccccccc}
		\hline
		& \multicolumn {14}{c}{Number-of-samples/speaker (Total speakers) } \\
		& \textit{10} & \textit{20} & \textit{30} & \textit{40} & \textit{50} & \textit{60} & \textit{70} & \textit{80} & \textit{90} & \textit{100} & \textit{110} & \textit{120} & \textit{130} & \textit{140} \\
		\textit{bits} & (15741) & (15741) & (15741) & (15740) & (15740) & (15061) & (13307) & (11524) & (9565) & (7601) & (5697) & (3943) & (2481) & (1363) \\
		\hline
		\hline
		\textit{1} & 42.90 & 34.52 & 31.67 & 30.22 & 29.34 & 28.71 & 28.23 & 27.87 & 27.58 & 27.32 & 27.15 & 26.97 & 26.85 & 26.77 \\
		\textit{2} & 87.78 & 64.18 & 56.01 & 51.90 & 49.42 & 47.71 & 46.44 & 45.49 & 44.74 & 44.09 & 43.63 & 43.16 & 42.83 & 42.63 \\
		\textit{3} & 142.67 & 96.42 & 80.03 & 71.63 & 66.50 & 62.98 & 60.37 & 58.41 & 56.85 & 55.56 & 54.58 &	53.66 &	52.96 &	52.47 \\
		\textit{4} & 225.51 & 144.74 & 113.96 & 97.97 & 88.20 & 81.55 & 76.68 & 73.02 & 70.12 & 67.74 & 65.88 &	64.20 &	62.86 &	61.82 \\
		\textit{5} & 332.24 & 216.51 & 166.78 & 139.37 & 122.15 & 110.33 & 101.69 & 95.15 & 89.99 & 85.79 & 82.41 & 79.46 & 77.03 & 75.05 \\
		\hline
	\end{tabular}
\end{table*}

\begin{table*}[!ht]
    \scriptsize
	\centering
	\caption{Uniqueness Estimates from VoxCeleb Dataset for Increasing Number of Samples}
	\label{tab:VoxCeleb2}
	\begin{tabular}{c|cccccccccc}
		\hline
		 & \multicolumn {10}{c}{Number-of-samples/speaker (Total speakers)} \\
		 & \textit{10} & \textit{20} & \textit{30} & \textit{40} & \textit{50} & \textit{60} & \textit{70} & \textit{80} & \textit{90} & \textit{100} \\
		 \textit{bits} & 6060 & 5611 & 5103 & 4645 & 4253 & 3872 & 3583 & 3294 & 3055 & 2830 \\
		\hline
		\hline
		\textit{1} & 43.99 & 33.17 & 29.34 & 27.44 & 26.19 & 25.26 & 24.61 & 24.19 & 23.81 & 23.53 \\
        \textit{2} & 86.19 & 61.01 & 51.55 & 46.68 & 43.56 & 41.32 & 39.76 & 38.66 & 37.72 & 37.00 \\
		\textit{3} & 133.67 & 89.37 & 72.91 & 64.36 & 58.93 & 55.07 & 52.33 & 50.34 & 48.68 & 47.36 \\
		\textit{4} & 203.53 & 129.79 & 101.50 & 86.71 & 77.42 & 70.93 & 66.29 & 62.85 & 60.05 & 57.80 \\
		\textit{5} & 279.70 & 178.92 & 137.29 & 114.83 & 100.63 & 90.76 & 83.64 & 78.30 & 73.99 & 70.53 \\
		\hline
	\end{tabular}
\end{table*}

\subsubsection{Number of Speakers} Capturing between-speaker variability essentially requires uniform sampling of i-vector space. 
This can be roughly achieved by using voice samples of a large number of randomly selected speakers. 
Table \ref{tab:80sample} provides estimates for increasing number of speakers when the number of samples for each speaker is fixed to 80, which sums up to 7 minutes of pure speech. 
These results overall show that estimates based on fewer speakers (10-30) result in a significant underestimation as distribution of i-vectors cannot be reliably obtained and $H({\bf V})$ is miscalculated.
At the other extreme, we observe that increasing the number of speakers by an order of magnitude, from one to ten thousand, the estimates change only marginally. 
Thus, it can be deduced that around a thousand speakers is sufficient to obtain accurate estimates as long as a large number of samples per speaker are available.
The table also incorporates the estimates obtained from the TED dataset in its last column for comparison, which includes around a thousand speakers with 71 samples per speaker on 
average \cite{tandogan2017towards}. 
Comparing the estimates obtained using the two datasets around the same number of speakers, a 1-4 bit difference is observed depending on the number of quantization bits. 
We believe this difference is potentially due to the fact that TED dataset includes more high quality speech samples than the TEDx dataset.

\begin{table*}[!ht]
    \scriptsize
	\centering
	\caption{Uniqueness Estimates from TEDx Dataset for Varying Number of Speakers (80 Samples/Speaker)}
	\label{tab:80sample}
	\begin{tabular}{c|ccccccc||c}
		\hline
		\multirow{2}{*}{\backslashbox{}{\textit{\# of speakers}} } & \multicolumn {7}{c||}{TEDx Dataset}  & TED dataset \cite{tandogan2017towards}\\
		 \textit{bits} & \textit{10} & \textit{30} & \textit{100} & \textit{300} & \textit{1000} & \textit{3000} & \textit{10000} & 921\\
		\hline
		\hline
		\textit{1} & 24.33 & 27.30 & 27.94 & 27.93 & 28.04 & 28.03 & 27.87 & 28.38 \\
        \textit{2} & 38.74 & 44.15 & 45.62 & 45.66 & 45.76 & 45.75 & 45.50 & 46.94 \\
		\textit{3} & 49.53 & 56.55 & 58.50 & 58.62 & 58.73 & 58.72 & 58.41 & 62.33 \\
		\textit{4} & 62.08 & 70.49 & 72.95 & 73.21 & 73.36 & 73.36 & 73.02 & 80.41 \\
		\textit{5} & 81.41 & 91.52 & 94.74 & 95.22 & 95.46 & 95.49 & 95.16 & 100.49 \\
		\hline
	\end{tabular}
\vspace{-0.5cm}
\end{table*}

\vspace{-0.3cm}

\textcolor{black}{
\subsubsection{Sample Duration} 
To determine how uniqueness estimates are affected from the length of speech samples, we repeated our measurements by also extracting i-vectors from speech samples with 
durations of 2.5 and 10 seconds.
To this objective, seventy 5-second long speech samples associated with 1,000 speakers in the measurement set of TEDx dataset are divided into two to obtain shorter speech samples. 
Similarly, consecutive speech samples extracted from each video are combined together to obtain longer ones. 
Table \ref{tab:TEDx-overall} provides resulting uniqueness estimates as well as the two terms involved in their calculation, {\em i.e.,} the entropy of population i-vectors, $H({\bf V})$, and average entropy of speakers' i-vectors, $H({\bf V}|S)$.
Essentially, we observe around 10-20 bits in increase in uniqueness estimates with every doubling of the duration.
Moreover, these results show that longer speech samples enable better speaker modeling as within-speaker variability, $H({\bf V}|S)$, decreases much faster than between-speaker variability, $H({\bf V})$, for longer duration samples.
This trend aligns with the change in speaker verification performance as well where EER values are found to be $2.8$, $1.1$, and $0.3$ for increasing durations.
}

\def\arraystretch{1.2}
\begin{table*}[!ht]
    \scriptsize
	\centering
	\caption{Uniqueness Estimates From TEDx Dataset for Varying Sample Durations  (70 Samples/Speaker)}
	\begin{tabular}{c|ccc|ccc|ccc}
		\hline
		     & \multicolumn{3}{|c|}{2.5 seconds} & \multicolumn{3}{|c|}{5 seconds} & \multicolumn{3}{|c}{10 seconds} \\
		{\textit{bits}} & $H({\bf V})$ & $H({\bf V}|S)$ & $I(S;{\bf V})$ & $H({\bf V})$ & $H({\bf V}|S)$ & $I(S;{\bf V})$ & $H({\bf V})$ & $H({\bf V}|S)$ & $I(S;{\bf V})$ \\
		\hline
		\hline
		\textit{1} & 198.82 & 180.32 & 18.50 & 198.63 & 170.87 & 27.76 & 197.90 & 157.08 & 40.82 \\
		\textit{2} & 390.77 & 359.81 & 30.96 & 367.97 & 322.29 & 45.68 & 331.22 & 265.59 & 65.64 \\
		\textit{3} & 579.27 & 536.56 & 42.71 & 543.42 & 483.94 & 59.48 & 497.52 & 415.17 & 82.35 \\
		\textit{4} & 767.35 & 706.66 & 60.69 & 726.09 & 650.39 & 75.70 & 677.54 & 580.09 & 97.45 \\
		\textit{5} & 943.04 & 851.44 & 91.59 & 896.71 & 796.09 & 100.62 & 845.41 & 728.03 & 117.38 \\
		\hline
	\end{tabular}
    \label{tab:TEDx-overall}
\end{table*}

\subsubsection{Diversity of Speech Samples}

The Movie Dialogues dataset contains a relatively small number of speakers and speech samples as compared to the TEDx dataset, but it provides a more diverse kind of utterances with wider emotional variability in voice.
Table \ref{tab:movie} presents corresponding results when speaker models are generated for an increasing number of samples per speaker.
It is immediately obvious that resulting uniqueness estimates are significantly lower than those found in Tables \ref{tab:samples-p-speaker} and \ref{tab:80sample}. 
This difference may partly be attributed to real-life and uncontrolled conditions of audio acquisition in movies; however, 
the quality of samples cannot be a significant factor in this as, ultimately, speech samples only include spoken words that could be matched to the subtitle. 
The major factor in play here is the increased within-speaker variability which induces ambiguity in the estimated speaker models. 
This is also supported by the earlier observation that application of audio effects on voice samples (such as changing loudness, shifting pitch, addition of background noise and echo) causes a somewhat similar reduction in estimates, though to a lesser extent \cite{tandogan2017towards}.

Overall, as the number of speakers with a large number of samples is limited, the degree of distinctiveness can be evaluated 
in conjunction with results of Tables \ref{tab:samples-p-speaker} and \ref{tab:80sample}.
As the results of Table \ref{tab:samples-p-speaker} demonstrate, using fewer speaker samples (10-30 samples per speaker) overestimates the inherent uniqueness. 
Hence, estimates should be expected to be lower than 18-52 bits, for 2-4 bit quantization.
At the same time, results of Table \ref{tab:80sample} indicate that estimates of uniqueness based on fewer speakers (30 or less) leads to an underestimation.
Therefore, estimates in the last three columns of Table \ref{tab:movie} can be interpreted as lower bounds. 
That is, at 2-4 bits quantization, estimates should be higher than 12-27 bits.
Therefore, based on measurements obtained using speech samples of 64-46 speakers with 60-70 samples per speaker, we estimate the individuality of human voice decreases to 13-31 bits level, 
under 2-4 bit quantization, 
when higher degree of within-speaker variability is taken into account.

\begin{table*}[!ht]
    \scriptsize
	\centering
	\caption{Uniqueness Estimates from Movie Dialogues Dataset for Increasing Number of Samples}
	\label{tab:movie}
	\begin{tabular}{c|cccccccccc}
		\hline
		 & \multicolumn {10}{c}{Number-of-samples/speaker (Total speakers)} \\
		 & \textit{10} & \textit{20} & \textit{30} & \textit{40} & \textit{50} & \textit{60} & \textit{70} & \textit{80} & \textit{90} & \textit{100} \\
		 \textit{bits} &556 & 286 & 184 & 132 & 93 & 64 & 46 & 33 & 26 & 21 \\
		\hline
		\hline
		\textit{1} & 14.66 & 10.48 & 8.95 & 8.08 & 7.53 & 7.26 & 7.00 & 6.77 & 6.78 & 6.92 \\
        \textit{2} & 34.62 & 22.38 & 17.94 & 15.90 & 14.52 & 13.59 & 12.88 & 12.26 & 12.13 & 12.30 \\
		\textit{3} & 64.45 & 40.20 & 31.19 & 27.03 & 24.03 & 21.78 & 20.05 & 18.62 & 18.09 & 17.88 \\
		\textit{4} & 113.56 & 69.40 & 52.56 & 44.77 & 39.13 & 34.90 & 31.45 & 28.60 & 27.28 & 26.41 \\
		\textit{5} & 184.27 & 116.28 & 88.12 & 74.49 & 64.59 & 56.6 & 50.43 & 45.25 & 42.45 & 40.50 \\
		\hline
	\end{tabular}
\vspace{-0.5cm}
\end{table*}

\vspace{-0.2cm}
\subsection{Comparison with Other Measures}

We also compared uniqueness estimates obtained through our mutual information based measure, \textcolor{black}{given in Table \ref{tab:TEDx-overall}}, with those generated using the relative entropy and Hamming distance based measures on the TEDx dataset as \textcolor{black}{presented} in Table \ref{tab:methods}. 
\textcolor{black}{To determine the degree of distinctiveness with respect to the measure introduced by Daugman \cite{IEEEexample:daugman2003},  we applied it to binary quantized, 200-dimensional i-vectors.
By measuring the degree of freedom in the distribution of pairwise Hamming distances between i-vectors, an entropy of 186.87 bits is measured.}
This result is in line with the interpretation that this measure estimates the number of independent components in the biometric feature vector. 

Similarly for the relative entropy based measure, we performed all steps described by Adler {\em et al.} in \cite{adler2009towards} to original, unquantized i-vectors obtained \textcolor{black}{from} 11,524 speakers, with 80 samples per speaker.
Accordingly, the relative entropy of i-vector distribution is determined to be $80.61$ bits.
As compared to values obtained on the TED dataset \cite{tandogan2017towards}, resulting estimates for both measures are found to be 10-30 bits lower.
This can essentially be explained by the more comprehensive nature of the TEDx dataset, which allows better incorporation of between-speaker variability due to an order of magnitude increase in the number of speakers.

\def\arraystretch{1.2}
\begin{table}[!ht]
	\centering
	\scriptsize
	\caption{Comparisons of Measures for Uniqueness Estimation }
	\begin{tabular}{c|cccc}
		\hline
		  & \textcolor{black}{Hamming} & Relative & \multicolumn{2}{c}{Mutual Inf.} \\
		  & \textcolor{black}{Distance} \cite{IEEEexample:daugman2003} & Entropy \cite{adler2009towards} & \textit{(2 bits)} & \textit{(3 bits)} \\
		\hline
		\hline
		  {TEDx Dataset}	& 186.87 & 80.61 & 44.82 & 56.80 \\
		  {TED Dataset \cite{tandogan2017towards}}	& 195.08 & 109.34 & 46.94 & 62.33 \\
		\hline
	\end{tabular}
	\label{tab:methods}
\end{table}

\vspace{-0.41cm}
\section{Discussion and Conclusions}
\label{sec:conclusion}

In this work, we seek to estimate the uniqueness of human voice with respect to the widely used i-vector representation of voice.
For this purpose, we introduce a mutual information-based measure for uniqueness estimation and evaluate it on custom and public datasets 
that more thoroughly capture \textcolor{black}{between-speaker} and within-speaker variability in the voice. 
The newly generated datasets include speech samples collected from public sources such as the TEDx Talks video archive and audio tracks of movies, 
and they are among the first of their kind used in such a study.
Most strikingly, measurements on several datasets show that quantization of i-vectors does not impair speaker verification performance measured in EER.
Our results show that uniqueness estimates obtained using speech samples of more than a thousand speakers with $100$ samples/speaker yields stable measurements, and
using a limited number of speakers and/or samples per speaker may result \textcolor{black}{in} significant deviation in the estimated values.
Findings also indicate that within-speaker variability is a more important factor affecting the reliability of estimates. 
It is observed that estimates drop significantly when speech samples are obtained under less controlled environments.  
In contrast, i-vectors extracted from longer speech samples allow for better speaker modeling and yield an increase in the uniqueness estimates.

\textcolor{black}{
Our biometric information measure can be applied to all biometric modalities that exhibit high within-subject variability. 
One limitation of our uniqueness measure in the context of using a feature representation of speech is that speaker embeddings in general cannot isolate speaker variability from channel and session related variations. 
Since large scale datasets, such as TEDx and VoxCeleb, include speech samples recorded under uncontrolled conditions, uniqueness estimates might be biased by some unaccounted characteristics.}
Another limitation of our work is that its individuality estimates are based on a feature representation determined by a generative model. 
As neural-network based speaker embeddings are becoming increasingly more prevalent, our measurements must also be expanded to discriminative models  
in order to more confidently estimate the degree of individuality of the human voice. 
We also note that the reliability of uniqueness estimates obtained on the new datasets depend on the accuracy of the text-to-speech alignment.
Errors in alignment may cause a false increase in the within-speaker variability and result in higher entropy measurements. 
This in turn may increase the EER and yield a decrease in the uniqueness estimates.

\vspace{-0.1cm}
\bibliographystyle{IEEEbib}
\bibliography{TIFS}
\onecolumn
\end{document}